# Dark Energy and Dark Matter as Inertial Effects

Serkan Zorba

Department of Physics and Astronomy, Whittier College

13406 Philadelphia Street, Whittier, CA 90608

szorba@whittier.edu

## ABSTRACT

A disk-shaped universe (encompassing the observable universe) rotating globally with an angular speed equal to the Hubble constant is postulated. It is shown that dark energy and dark matter are cosmic inertial effects resulting from such a cosmic rotation, corresponding to centrifugal (dark energy), and a combination of centrifugal and the Coriolis forces (dark matter), respectively. The physics and the cosmological and galactic parameters obtained from the model closely match those attributed to dark energy and dark matter in the standard Λ-CDM model.

# Introduction

The two most outstanding unsolved problems of modern cosmology today are the problems of dark energy and dark matter. Together these two problems imply that about a whopping 96% of the energy content of the universe is simply unaccounted for within the reigning paradigm of modern cosmology.

The dark energy problem has been around only for about two decades, while the dark matter problem has gone unsolved for about 90 years. Various ideas have been put forward, including some fantastic ones such as the presence of ghostly fields and particles. Some ideas even suggest the breakdown of the standard Newton-Einstein gravity for the relevant scales.

Although some progress has been made, particularly in the area of dark matter with the nonstandard gravity theories, the problems still stand unresolved.

In the standard model of cosmology, the cosmic inflation, dark energy and dark matter are all different patches artificially sewn to the Big Bang model in an effort to address various unsettling observations about our universe. Why did the universe undergo an accelerated expansion in the very early moments of its existence, then subside for a long time, and recently decide to start accelerating again? Why these intermittent, and confusing, acceleration epochs?

What is behind these accelerations? Why are there anisotropies in the cosmic microwave background radiation, as revealed by WMAP and Planck satellite data, pointing to a cosmic axis? And what exactly is dark matter that even our most sophisticated detectors cannot detect? These are some of the formidable questions of modern physics and cosmology that the Big Bang model is increasingly failing to answer satisfactorily.

In this article, I propose an alternative model, a globally rotating (not a solar-system type rotation) disk-shaped spacetime as part of a larger universe to address these problems in our smaller universe. My motivation is simple, when we observe mysterious effects in our observable universe, whose supposed causes we cannot seem to be detecting, our first reaction should not be to hypothesize undetectable ghostly particles or modify our standard and normally very successful laws of physics. Rather, the common sense would require that we first explore whether we are living in a non-inertial universe. This is especially so since the properties of dark energy and dark matter are so "mysteriously" gravity-like. What better gravity-like effects than inertial effects such as the good old centrifugal and Coriolis forces? Indeed, the equivalence principle teaches us the equivalence of gravitational and inertial effects.

My humble calculations in this model do produce—in one stroke and, for me, in a fine display of *unification*—the main parameters of both dark energy and dark matter with a good degree of precision, and thus give a strong support for the rotary paradigm to resolve the two most outstanding problems of modern cosmology. In other words, this model does naturally *unify* these two seemingly unrelated phenomena.

The obvious points of opposition to the disk-shaped rotating model are expressed as follows:

(a) The universe is rotating with respect to what?

(b) It seems to be patently at odds with the isotropy aspect of the cosmological principle.

My response to these points are:

The rotation of the spacetime of our universe implies that there is extraneous physics, i.e., our universe is not a closed system, and that there is a larger universe or cosmically parallel universe(s). And the implied cosmic axis of the model must have had axial precession (see Fig. 1). The resulting overall universe would be disk-shaped, but if the size of our observable universe is only a fraction of the overall, but anisotropic, universe, then we will mistakenly assume cosmic isotropy (see Fig. 2). Furthermore, my model posits a cosmic rotation with an angular speed that is currently equal to the Hubble constant, and so it is a bit less than $10^{-10}\ rad/year$, which is very very slow. Our current observational techniques cannot yet detect such a small global rotation. However, there are observations forcing us to question the Big Bang model, in addition to the problems of dark energy and dark matter, e.g., the so-called cosmic "axis of evil," potentially supporting my hypothesis.

The article consists of two parts. Part I will focus on the implications of the rotary model on the dark energy problem. Part II will be on the implications of the rotary model on the dark matter problem.

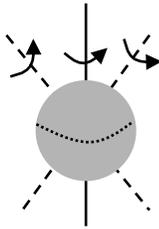

FIG. 1. The cosmic axis must have had an axial precession.

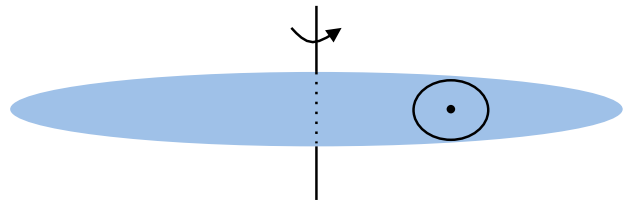

FIG. 2. If our observable universe is a part of a larger anisotropic, say, disk-shaped, universe, then we will mistakenly assume cosmic isotropy.

## Part I: Dark Energy as an Inertial Effect of Cosmic Rotation

One of the outstanding problems in cosmology, alas in all of physics, is the problem of the recent acceleration of our universe as evidenced by the astronomical observations of the last two decades.[1-4] The acceleration is thought to be caused by some inscrutable energy termed dark energy. Many ideas have been put forward as to the nature of dark energy: cosmological constant, phantom energy, quintessence, etc.[5-7] The cosmological constant is the best and simplest of these candidates to explain dark energy.[8] Unlike phantom energy and quintessence, the cosmological constant is presumed to be static and homogeneous.

Unraveling the underlying physics and nature of the cosmological constant/dark energy, however, has been proving very challenging, if not embarrassing, for physicists.[8-10] To fully understand the nature of the cosmological constant/dark energy, as aptly explained by cosmologist Sean Carroll, one has to confront three formidable problems:[8]

1. The extreme smallness of the cosmological constant compared with what is expected from quantum field theoretic calculations, and the lack of a physical basis from which we can calculate it.

2. The ultra smoothness of the dark energy associated with the cosmological constant.

3. The approximate equality of the dark energy density to the matter density in today's cosmic era (the so-called cosmic coincidence scandal), and the recent onset of cosmic acceleration.

It is clear that the success and acceptance of a cosmological model put forward to explain the problem of the accelerating universe and cosmological constant/dark energy will depend on how well it addresses the above three problems.

In a recent work,[11] I have postulated a rotational cosmic model, whereby the universe is posited to be rotating globally but very slowly with an angular speed equal to the Hubble constant $\omega = H \cong 2.2{\times}10^{-18}\ rad/sec$. This model provides a natural and simple explanation to the problem of accelerating universe and obviates many a problem associated with it.

In this first part of the article, I will show how the rotational model of the universe offers a viable solution to each and every one of the foregoing three problems associated with dark energy, and how the main physical results/parameters obtained from the rotary model match exactly those attributed to vacuum/dark energy in the concordance model.

The standard model of cosmology today presumes an irrotational, homogenous and isotropic universe. However, the presence of a slow rotation cannot be ruled out, inasmuch as its deviation from a homogenous and isotropic universe would be minimal and escape detection. Moreover, we are compelled to investigate other potential models since there are major unsolved problems in the standard model of cosmology waiting for answers, such as the cosmological constant/dark energy problem, dark matter problem, horizon problem, and flatness problem.

A rotational universe is perfectly viable with General Relativity (GR) as shown first by Gödel.[12] In fact, it is common to study rotational universe models, usually within the framework of Bianchi cosmologies, but not all.[13]

To facilely elicit the physical effects of global rotation of a disk-shaped universe on non-inertial as well as coordinate (inertial) observers, I choose to work in cylindrical coordinate system: picture a platform rotating with a constant angular frequency $\omega$. The line element will be given by[14] (from $ds^2 = g_{\mu\nu}dx^\mu dx^\nu;\ \mu, \nu = 1,2,3,4$)

$$ds^2 = dr^2 + r^2 d\theta^2 + dz^2 + 2r^2 \frac{\omega}{c} d\theta d\tau - \left(1 - r^2 \frac{\omega^2}{c^2}\right) d\tau^2. \tag{1}$$

The corresponding nonzero covariant metric components and Christoffel symbols are as follows

$$g_{11} = 1\ ;\ g_{22} = r^2\ ;\ g_{33} = 1\ ;\ g_{44} = -1 + \frac{r^2\omega^2}{c^2}\ ;\ g_{24} = g_{42} = r^2 \frac{\omega}{c}.$$

$$\Gamma^1_{22} = -r\ ;\ \Gamma^1_{24} = \Gamma^1_{42} = -r\ \omega/c\ ;\ \Gamma^1_{44} = -r\ \omega^2/c^2\ ;\ \Gamma^2_{12} = \Gamma^2_{21} = 1/r\ ;\ \Gamma^2_{14} = \Gamma^2_{41} = \omega/rc.$$

The conserved (covariant) energy component, as observed by an observer on the rotating platform, can be shown to be

$$E_{coord} = p_4 = g_{24}p^2 + g_{44}p^4 = mc^2\left(1 - \frac{r^2\omega^2}{c^2}\right) = E_{inertial}\left(1 - \frac{r^2\omega^2}{c^2}\right). \tag{2}$$

For our model of a spinning universe, the energy component of the covariant momentum four-vector will be conserved as an object moves over a geodesic path because our metric has a very weak time dependence. $E_{inertial}$ is the matter-energy content with respect to an inertial observer and is given in terms of the rest mass energy, $E_0$, as

$$E_{inertial} = E_0\left(1 - \frac{r^2\omega^2}{c^2}\right)^{-1/2} = E_0/\sqrt{1 - \xi^2}, \tag{3}$$

where $\xi^2 = \frac{r^2\omega^2}{c^2}$ is used for short hand. The maximum value $\xi$ can take is 1, which corresponds to the Hubble radius. (Note that the Hubble sphere constitutes our gravitational horizon, as explained recently by Melia and coworkers.[15])

Using our covariant metric components, we obtain the covariant energy as observed by the coordinate observer at the axis of rotation

$$E_{coord} = p_4 = g_{\mu\nu}p^\nu = E_{inertial}(1 - \xi^2) = E_0\sqrt{1 - \xi^2}. \tag{4}$$

Therefore, the magnitude of the energy expression due to rotation is the difference between the matter-energy content observed by the inertial observer and that of the rotational (coordinate) observer.

$$|E_{rot}| = E_{inertal} - E_{coord} = E_0\left((1 - \xi^2)^{-1/2} - \sqrt{1 - \xi^2}\right) = \frac{E_0\xi^2}{\sqrt{1-\xi^2}}. \tag{5}$$

This is the energy expression whose average value the coordinate observer will overlook when he/she surveys the rotating universe. The average values of these energy functions on the interval $[0, 1]$ are found as follows (see Fig. 3 below)

$$\langle E_{inertial}\rangle = \int_0^1 E_0/\sqrt{1 - \xi^2}\,d\xi = E_0\pi/2. \tag{6}$$

$$\langle E_{coord}\rangle = \int_0^1 E_0\sqrt{1 - \xi^2}\,d\xi = E_0\pi/4. \tag{7}$$

$$|\langle E_{rot}\rangle| = \langle E_{inertial}\rangle - \langle E_{coord}\rangle = \langle E_{coord}\rangle = E_0\pi/4. \tag{8}$$

Hence, the coordinate observer will overlook an average matter energy that corresponds to the average energy of the rotation of the universe, and which is at the same time identically equal to the average matter-energy content of the coordinate observer's own universe. In general, note from Fig. 3, that the overlooked dark energy (the difference between the energies of the inertial and coordinate observers) is small—compared with the energy content of the coordinate observer's universe—for the early ages of the universe, and then catches up, and finally for today's age corresponding to the Hubble radius it becomes much larger. The latter statement rings a familiar tone especially with the large "missing" energy content of our universe today.

Consequently, in a rotating universe, there will be some extra energy that will be overlooked by the coordinate observer. One must take that extra energy into account much in the same way the dwellers of the Earth take into consideration the inertial forces (centrifugal and Coriolis effects) to properly explain their observations of, for instance, large scale weather phenomena.

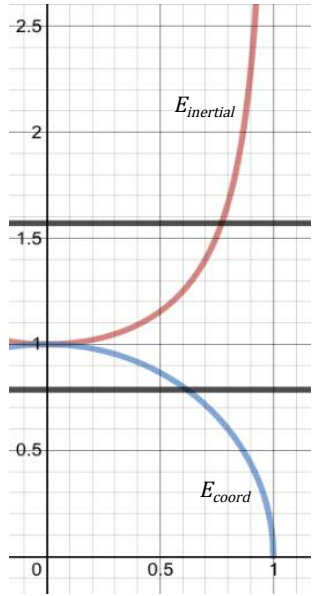

FIG. 3. Matter-energy content of a rotating universe (expressed as a ratio over the rest-mass energy) is plotted as a function of reduced distance $\xi$ as observed by an inertial observer (top curve), and a coordinate observer (bottom curve). The horizontal black lines represent the average values of these respective functions. We note that the average energy observed by the inertial observer is exactly twice that observed by the coordinate observer. The difference is due to rotation. The energy overlooked by the coordinate observer is identically equal to the observed energy content of his/her universe.

Thus the absolute total energy of a rotating universe, expressed in terms of the matter-energy density as observed in the rotating system, must be that which is observed by the inertial observer. That means, on the average, $\rho_{tot} = 2\rho_{rot} = 2\rho$.

Therefore, on the average, the overlooked energy is going to be of the same order of magnitude as the matter density. Astronomical measurements inform us that today the energy density due to dark energy is of the same order of magnitude as that due to matter density.[16] Indeed, as stated earlier, the former fact is part of the cosmic coincidence problem.[17] The rotary model thus provides a physical basis for the cosmological constant/dark energy. The model also shows why it is so small—as compared with the quantum vacuum energy calculations. It is because dark energy is not due to the putative vacuum energy, but rather due to a cosmic rotation, which is very slow, and the angular frequency of the rotation is the Hubble constant.

What is the effect of rotation in a quasi-Newtonian treatment, as is commonly employed in many a textbook on GR?[18-20] Consider a sphere of mass $M$ and radius $r$, which is rotating around some axis with an extremely small angular frequency $\omega$. Now imagine a galaxy of mass $m$ at the surface of that sphere. In the reference frame of the galaxy, the equation of motion will be given by the following expression

$$m\ddot{r} = -\frac{GmM}{r^2} + m\omega^2 r, \tag{9}$$

where the first term on the RHS is gravitational attraction and the second term is the fictitious (very real for the galaxy) outward-pulling centrifugal force. Integrating this equation produces

$$m\frac{\dot{r}^2}{2} = +\frac{GmM}{r} + \frac{m\omega^2 r^2}{2} + C, \tag{10}$$

where $C$ is the integration constant, which corresponds to the total energy content of the sphere. In other words, $C$ is nothing but the total energy of the system and can be written as

$$E_{tot} = E_k + V + \phi_{rot} = m\frac{\dot{r}^2}{2} - \frac{GmM}{r} - \frac{m\omega^2 r^2}{2} = -\frac{kc^2 mr_0^2}{2R_0^2}, \tag{11}$$

where $E_{tot}$ is written in terms of the time-independent curvature parameter, $k$, of the universe, and $R_0$ the radius of the universe at the current epoch.[20] See Fig. 4 for a comparison of the gravitational and centrifugal-force potential energy functions.

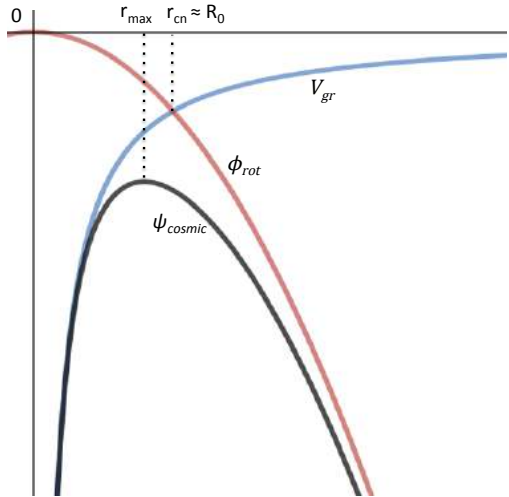

FIG. 4. (Quasi-Newtonian) Rotational potential, $\phi_{rot}$, and gravitational potential, $V_{gr}$, energies of the rotary model are co-plotted versus distance. Note how the strengths of the two energies coincide in almost today's era, represented by the Hubble radius $R_0$. The resulting cosmic potential $\psi_{cosmic}$ is also shown. We see that the universe rolled to the potential maximum (zero net-force) recently, beyond which universal acceleration starts. Today we are more or less at the cosmic potential hilltop. From there on our cosmic future will be all "downhill."

In an expanding universe, $r = r_0 a(t)$, where $r_0$ is the radius of our sphere at some arbitrary time, and $a(t)$ is a time-dependent scale factor. Eq. (11) can now be written explicitly as

$$\frac{1}{2} m r_0^2 (\dot{a})^2 - \frac{4\pi m G r_0^2 a^2 \rho}{3} - \frac{1}{2} m (r_0 a)^2 \omega^2 = -\frac{k c^2 m r_0^2}{2 R_0^2}, \tag{12}$$

where $\omega$ is the angular frequency of the rotation - and according to my spinning universe model it is equal to the Hubble constant H at the present time - $G$ is the gravitational constant, $\rho$ is the average mass-energy density of matter in the universe, and $c$ is the speed of light in vacuum. Rearranging Eq. (12), we get the Friedmann equation for my model as

$$(\dot{a})^2 = \frac{8\pi G \rho a^2}{3} - \frac{k c^2}{R_0^2} + a^2 \omega^2. \tag{13}$$

Written in terms of the Hubble parameter, Eq. (13) becomes

$$H^2 = \frac{8\pi G \rho}{3} - \frac{k c^2}{R_0^2 a^2} + \omega^2. \tag{14}$$

Since the standard Friedmann equation, with the cosmological constant, is given as

$$H^2 = \frac{8\pi G \rho}{3} - \frac{k c^2}{R_0^2 a^2} + \frac{c^2}{3} \Lambda, \tag{15}$$

where $\Lambda$ is the cosmological constant,[21] $\Lambda = 3\omega^2/c^2$. Further, since $\omega = H$ in my model, $\Lambda = 3H^2/c^2$. This is very significant, because like the relativistic treatment, the quasi-Newtonian treatment of a rotational universe also naturally produces an additional energy term, which in this case corresponds exactly to the cosmological constant term.

To understand the nature of the forces present during rotation, we need the geodesic equation, $a^i = -\Gamma_{\mu\nu}^i v^\mu v^\nu + (v^i/c)\Gamma_{\mu\nu}^4 v^\mu v^\nu;\ i = 1, 2, 3$. One can calculate the radial and azimuthal accelerations to be $a_r = r\omega^2 + 2\omega r\dot{\theta}$ and $a_\theta = -2\omega\dot{r}$, respectively. For an object with no or negligible azimuthal motion, one obtains $a_r = r\omega^2 = v^2/r$. So $v = r\omega$, which is equivalent to Hubble's law. (So we see that the constant angular velocity assumed in my model universe must indeed be the Hubble constant. This means that my model places an observational constraint of about $10^{-10}$ rad yr$^{-1}$ on the rotation of our universe. This compares well with a recent estimate.[12]) The corresponding radial force will be the centrifugal force given by

$$F_r = m r \omega^2; \tag{16}$$

For a radially moving galaxy, the galaxy observer will observe not only an inward-pushing gravitational attraction, but also an outward-pulling tension-like (anti-gravitational) centrifugal force due to rotation. Furthermore, since $\nabla \times F_r = 0$, the centrifugal force will act as a conservative force. Thus the galaxy observer will associate with it a potential (dark) energy of the form $\phi_{rot} = -\int F_r \cdot dr = -m\omega^2 r^2/2$. The latter can also be looked upon as a type of negative kinetic energy, as employed in phantom field models of dark energy.[7,8]

Here, we should also note that the rotary model provides the explicit form of the "dark force," corresponding to the dark energy, to be $F_r = mH^2r$, which would be significant only for very large distances of the order of the size of our universe. This is because $H^2 = \sim 10^{-36}sec^{-2}$ is very small. This accords well with the standard model's expectation of the force responsible for dark energy, which is given by $F_{dark} = \lambda\, r$, where $\lambda$ is a very small number, which would make $F_{dark}$ important only for very large distances on the order of the size of the universe.[22]

Now returning back to our relativistic energy treatment, let us compare the rotational energy $E_{rot}$ with the gravitational potential energy $V_{gr}$ (see Fig. 5). However, one has to note that from the perspective of the coordinate observer the rotational energy will appear as a "dark" potential energy since the objects will appear to be hurled away from the observer as if with a centrifugal force. This is as if the rotational potential energy, as inferred by the coordinate observer, dilutes the total absolute (inertial) energy and leaves behind only an apparent (coordinate) energy. That is, $E_{coord} = E_{inertial} + E_{rot}$. This implies that $E_{rot} = -\frac{E_0\xi^2}{\sqrt{1-\xi^2}}$.

From Fig. 5, we see that the rotational potential energy, $E_{rot}$, creates a force to pull an object to infinity, while the gravitational potential energy, $V_{gr}$, creates a force to push the object to the origin. They thus are in a tug of war, in the early stages (after the superluminal rotation, causing inflation, subsided) of which the gravitational attraction is dominant and decelerates the expansion, and in the later stages of which the rotational potential energy dominates and accelerates the expansion, as the distance gets larger. The resulting cosmic potential is also shown in Fig. 5. It first rolls up to a maximum (before which, the net force is negative), and then rolls downhill (the net force is positive). The plateau is where the corresponding forces balance out, beyond which the universal acceleration takes off. My rotary model thus captures exactly what the Type Ia supernova measurements of Perlmutter, Schmidt and Riess have revealed about the history of our universe about two decades ago.

It is therefore an observational fact that the universe has recently started to accelerate. However, it is not known why. In the words of Anderson and coworkers in a recent paper "Explaining the late-time acceleration of the expansion rate of the Universe is one of the most perplexing problems in modern physics."[23] Recent observations pin down the time of onset between 5 and 6 billion years ago. This corresponds to a redshift of about $z_{obs.\ onset} = 0.57$.[24]

Calculating, in my model, where the two forces balance out, we find the value of the cosmic radius at which the two opposing forces counterbalance each other (the plateau in Fig. 5) to be $r_{coin.} = 0.65R_0$. This gives a redshift of $z_{rot.\ model\ onset} = 0.54$ for the onset of the acceleration of the universe. Accordingly, the constraint my model places on the turning-point of cosmic acceleration is stunningly close to the observational value quoted above.

Let us now look at the evolution of the energy density corresponding to the rotational potential (dark) energy to see how smooth it is. Recalling the conversion factor between mass and length scales $\frac{G}{c^2} \times mass = length$,[25] and $\omega = H$,

$$\rho_{rot} = \frac{|E_{rot}|/c^2}{Volume} \sim \frac{m\xi^2}{\sqrt{1-\xi^2}} \frac{1}{r^3} = \frac{H^2}{G\sqrt{1-r^2H^2/c^2}} \sim a^0, \tag{17}$$

$$\rho_M = \frac{V_{gr}}{Volume} \sim \frac{M/r}{r^3} \sim \frac{1}{r^3} \sim a^{-3}, \tag{18}$$

where $a$ is the scale factor as used in Eq. (12). The rotational energy density $\rho_{rot} \sim 10^{-24}/\sqrt{1-10^{-52}r^2}$ is an extremely smoothly varying function of distance $r$, almost constant throughout most of spacetime, except for distances very close to the Hubble radius. On the other hand, the matter density $\rho_M$ falls off as $a^{-3}$. Therefore, the rotational energy density of the postulated model is a very smoothly-distributed density, as is the case for the so-called dark/vacuum energy.[8] Furthermore, comparing Eq. (17) with $\rho \propto a^{-3(1+w)}$, we deduce that $w = -1$ for our rotary model.

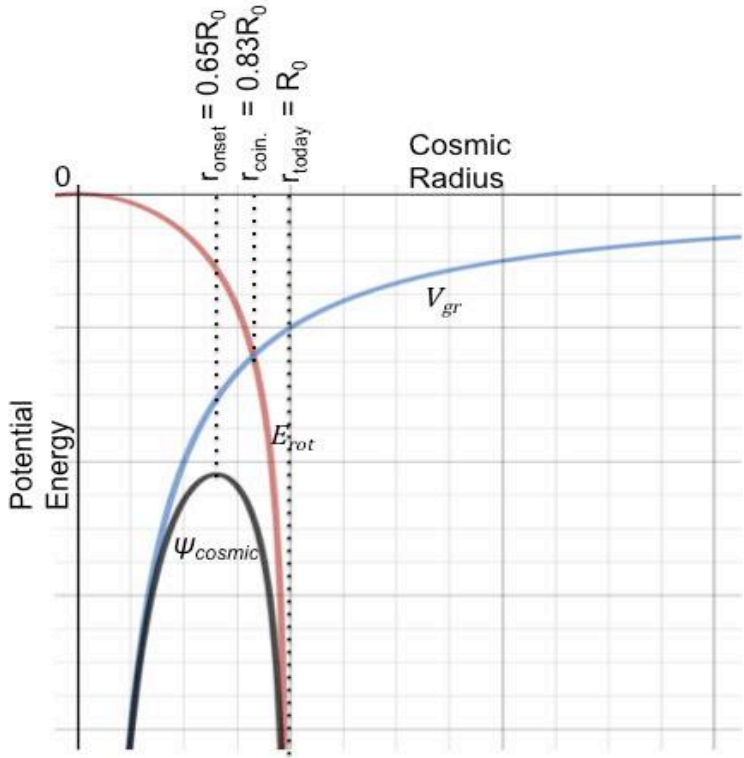

FIG. 5. (Relativistic) Rotational potential, $E_{rot}$, and gravitational potential, $V_{gr}$, energies of the rotary model are co-plotted versus cosmic distance. Note how the strengths of the two energies coincide in almost today's era, represented by the Hubble radius $R_0$. The resulting cosmic potential $\psi_{cosmic}$ is also shown. We see that the universe rolled to the potential maximum (zero net-force) recently, beyond which universal acceleration starts. Today we are more or less at the cosmic potential hilltop. From there on our cosmic future will be all "downhill."

Finally, let us see how the rotational model fairs vis-à-vis the so-called cosmic coincidence scandal, i.e., the comparable magnitudes of the dark energy and matter energy densities today. To see at what distance scale the two energies will balance out, we solve for the intersection point in Fig. 5.

$$-\frac{mr^2\omega^2}{\sqrt{1-\frac{r^2\omega^2}{c^2}}} = -G\frac{mM}{r}. \tag{19}$$

Solving the ensuing cubic equation, one finds the only real root to be (again, using the conversion factor between mass and length scales),

$$r_{coin.} = 0.83R_0. \tag{20}$$

This corresponds to a redshift parameter of $z = 0.20$, which is very close to today's cosmic age ($z = 0$). Thus the presented simple model also naturally reproduces the cosmic coincidence, clearly showing how the energy values of the two cosmic players more or less coincide in this cosmic age.

Summarizing the implications of Fig. 5, we see that (after the inflation period) the gravitational attraction was much stronger in the early universe. Then the rotational dark (centrifugal) force took over about 5 billion years ago ($z\sim0.54$), marking the onset of cosmic acceleration. The latter will dominate the gravitational force in the late universe. All this accords remarkably well with the expectations of the $\Lambda$-CDM model, and provides a solid answer to the fine-tuning problem associated with the cosmic coincidence.[26]

Moreover, since a de Sitter universe, (an empty universe model with $\rho = k = p = 0$, which is the working model for many inflationary cosmologies) has only a positive cosmological constant to be the dominant term in the Friedmann equation, it is clear that the de Sitter model is nothing but a holding-onto the rotational energy of the spacetime of our universe, while stripping it out of its material content, which results in a scale factor expansion as $a(t) \propto e^{\omega t} = e^{Ht}$.

Incidentally, a similar exponential expansion also ensues classically in a situation where the centrifugal force—with which I am associating the effect of a cosmological constant, $\Lambda$, in a de Sitter space—is the only dominant perceived force, as in a centrifuge. In the frame of reference of the object feeling only the centrifugal force, Newton's second law can be written as $\ddot{r} - \omega^2 r = 0$, which gives rise to a non-trivial solution of $r = r_0 e^{+\omega t}$, with $\omega = H \propto \sqrt{\Lambda}$. The classical time-dependent scale factor here can be seen to be $e^{Ht}$.

The model that I propound here would thus seamlessly incorporate not only dark energy and dark matter, but also the cosmic inflation into its physics, and answer the most outstanding questions of cosmology today as mentioned at the beginning. For example, take the intermittent nature of the accelerated expansion. Initially, the angular speed of the universe (what I call the Big Spin) was immense. The superluminal rotation frustrated gravity, giving rise to a brief but exponential expansion. As the universe expanded, the rotation speed slowed down to below the speed of light c, and gravity regained power and slowed the expansion (but not stopped it). Finally, with the ongoing expansion, the universe got bigger, and although the rotation got slower, the radius of the universe became big enough that the centrifugal force started to dominate at large distances over gravity.

Another surprise of the Big Spin model is to potentially explain the origin of the intrinsic quantum spin and the meaning of Planck's constant. For that, we will apply the conservation of angular momentum to the initial and current moments of the universe to make an estimate. The total number of quantum degrees of freedom in the universe is estimated by many authors to be around $10^{122}$. This number is part of a well-known cosmological large-number coincidence.[27]

With this in mind, we can setup the following conservation of momentum equation:

$I * \omega \cong 10^{122} * S$, where $I$ is the cosmic moment of inertia, $\omega$ the angular speed (which is Hubble's constant), and $S$ the angular momentum imparted to each quantum degree of freedom. This equation can be also written as $MR^2 * H \cong 10^{122} * S$. Using the most current values for these parameters, we obtain $S \approx 10^{-34} J.s.$, which is, remarkably, the order of magnitude value of the Planck constant. This argument can be turned the other way around as well. That is, one can calculate the total number of quantum degrees of freedom of the universe (the cosmological large-number coincidence) by dividing the total rotational energy of the universe (which we showed to be the same order of magnitude as the mass-energy content of the universe) by what I will call the "Hubble quantum" and define as $\hbar H = 2.3 \times 10^{-52} J$. Thus we will get $\frac{Mc^2}{\hbar H} \cong 10^{121}$. This thesis about the origin of the intrinsic quantum spin going back to the initial moments of the universe might prove useful toward achieving the holy grail of modern physics, that is, a viable theory of quantum gravity.

The rotary model presented herein thus addresses and answers more than the three issues associated with the cosmological constant/dark energy problem, as outlined at the beginning, in a satisfactory manner. For a convenient comparison, the predictions of the rotary universe model and those of the observation-based standard model are tabulated in Table I. The agreement between the two is stunning, and, in my opinion, cannot simply be a result of happenstance.

**TABLE I. 19 reasons as to why it was a Big Spin, not a Big Bang.**

| Physics | Observation/Standard Model | Rotary Universe Model ($\omega = H$) |
|---|---|---|
| 1- Hubble's law | $v = rH$<br>No insight. | $v = r\omega$<br>Natural result of rotation of spacetime. |
| 2- Small cosmological Constant | $\Lambda = 3\frac{H^2}{c^2}$<br>Artificially added to the EFE, no physical basis. | $\Lambda = 3\frac{\omega^2}{c^2}$<br>Rotation produces it, in GR and quasi-Newtonian treatments. |
| 3- Force | $F_\Lambda \sim + \lambda r$,<br>Repulsive; tension-like; significant only for very large distances. | $F = +m\omega^2 r$<br>Repulsive; tension-like; significant only for very large distances. |
| 4- Vacuum energy density | $\rho_\Lambda \sim a^0$ (Smooth) | $\rho_{\rm rot} \sim a^0$ (Smooth) |
| 5- Coincidence scandal | $\rho_\Lambda \sim \rho_{\rm M}$ (today) | $\rho_{\rm rot} \sim \rho_{\rm M}$ (today) |
| 6- Onset of acceleration | $z \sim 0.57$ | $z \sim 0.54$ |
| 7- Past strength | Gravity $\gg$ Dark energy | Gravity $\gg$ Rotational energy |
| 8- Today's strength | Gravity $\sim$ Dark energy | Gravity $\sim$ Rotational energy |
| 9- Future strength | Dark energy $\gg$ Gravity | Rotational energy $\gg$ Gravity |
| 10- Cosmic axis | The standard model is clueless. But the WMAP and Planck satellite data evince a preferred direction dubbed the cosmic "axis of evil." | A rotation axis is inherent in the model. |
| 11- Cosmic inflation<br><br>&Why intermittent accelerated epochs? | De Sitter expansion of the form $e^{Ht}$.<br>Nothing to say. | Centrifugal force expansion of the form $e^{\omega t}$.<br><br>Initial: Superluminal rotation frustrated gravity, hence accelerated expansion.<br>Intermediate: As the universe expanded rotation speed slowed down to below c, gravity regained power and slowed the expansion<br>Final: As the universe got bigger, although rotation got slower, the radius of the universe got bigger and hence centrifugal force dominates at large distances over gravity. |
| 12- Origin of the quantum Spin and Planck's constant | Nothing to say. | Proposition: The Big Spin imparted the inherent spin to all elementary particles at once…Planck's constant finds meaning. |
| 13- Dark matter acceleration mystery | Mysterious acceleration constant: $2\pi \times 10^{-10}\,m/s^2$ | Critical acceleration parameter:<br>$Hc = 6.59 \times 10^{-10}\,m/s^2$ |
| 14- Dark matter potentials | Linear and quadratic potentials presumably due to "dark matter." | Linear and quadratic potentials due to cosmic Coriolis and centrifugal forces. |
| 15- About galactic center | Gravity dominates. | Gravity dominates. |
| 16- About galactic edges | Linear term dominates. | Linear Coriolis term dominates. |
| 17- Beyond galactic disk | Significant quadratic term. | Significant quadratic centrifugal term. |
| 18- Relation between dark energy and dark matter | No relation. | Both are naturally unified in one effect. |
| 19- Galaxy spin and hurricane-like shape | No *satisfactory* explanation. | Natural result of the Big Spin, and an interplay between the ensuing inertial effects and gravity. |

## Part II: Dark Matter as an Inertial Effect of Cosmic Rotation

The problem of galaxy rotation-curve anomalies (RC) is the second most vexatious unsolved problem of modern cosmology, the other being the problem of the recent onset of acceleration in the expansion of the universe.[4,23,28]

Two leading hypotheses to explain the RC anomalies are dark matter (DM) scheme and modified gravity scheme.[28,29] The former posits the existence of some nonbaryonic matter, which only gravitationally interacts with the baryonic matter (and perhaps through the weak force). The latter contends that the standard Newton-Einstein gravity theory should be modified, especially for small accelerations.

Most astronomers' expectations lie with DM[30], albeit no independent observation of dark matter has ever been successful so far.[28] Although initially viewed with suspicion, the modified gravity paradigm has recently been garnering more support from observations.[31] Among the latter are the Modified Newtonian Dynamics (MOND), the conformal gravity (CG) and the Metric Skew Tensor Gravity (MSTG) theories.

In spite of the above-mentioned distinction between the two schemes, the whole problem of RC anomalies is usually referred to as the "dark matter" problem, perhaps to convey the idea that the problem is still far from being solved.

In this part of the article, I advance the claim that the DM problem is also a result of an inertial effect due to the postulated cosmic rotation around a cosmic axis, corresponding to a conjunction of cosmic Coriolis and centrifugal forces.

From the geodesic equation invoked in Part I, the involved forces will be the centrifugal and Coriolis forces, given respectively by

$$F_r = mr\omega^2; \; F_\theta = -2m\omega\dot{r}. \tag{21}$$

Consequently, in addition to the usual gravitational force the constituents of a galaxy experience (as an attraction to the galactic center), there will be two extra contributions coming from the cosmic rotation: a cosmic centrifugal force and a cosmic Coriolis force. Fig. 6 sketches the corresponding acceleration due to the former on several points on a hypothetical circular galaxy, whose axis of rotation is assumed to be parallel to that of the postulated cosmic rotation (also in Fig. 7). Note that for those cases where the two axes are not initially parallel, they would become parallel soon by virtue of the conservation of angular momentum.

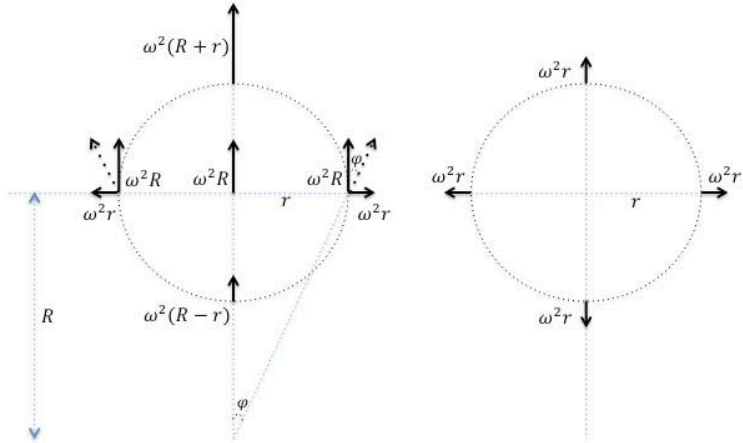

FIG. 6. (Left) Cartoon of cosmic centrifugal accelerations at various points along a circular galaxy, which is at a distance R from the center of rotation shown at the bottom. (Right) The resultant centrifugal accelerations from the perspective of the center of mass of the galaxy. Note that we assume $r \ll R$, and the drawings are not to scale. The corresponding forces are antigravity-like, tending to pull matter away from the galactic center. Note that the only parameters appearing in the centrifugal acceleration expression are the angular frequency of cosmic rotation and the distance of a particle to the galactic center. Hence, like the gravitational force (not shown), the centrifugal force is also uniform for a given distance from the galactic center.

Using the centrifugal force equation in Eq. 21, one can show (see Fig. 6) that the resultant centrifugal force is anti-gravitational. It tends to pull particles away from the galactic center of mass. However, like the gravitational force, it is uniform for a given distance from the galaxy center. Note that in Fig. 6 capital R is used for the cosmic radius, and small r is used for the galactic radius.

Let us add the cosmic Coriolis force term to the picture. Using the second equation in Eq. 21 with Hubble's law $\dot{r} = v = \omega R$, one finds (see Fig. 7 left panel) that the Coriolis force is, like gravity, centripetal, although non-uniform for a given galactic radius. The latter is because, unlike the cosmic centrifugal force, the cosmic radius does not disappear from the Coriolis term.

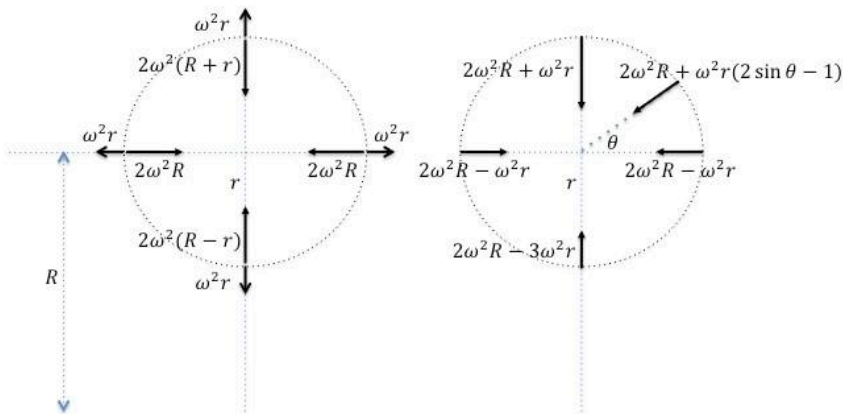

FIG. 7. (Left) Cartoon of both cosmic Coriolis and centrifugal accelerations at various points along a circular galaxy, which is at a distance R from the center of rotation shown at the bottom. Note that unlike the repulsive centrifugal force, the Coriolis force is attractive and changes along the orbit. (Right) The resultant cosmic accelerations from the perspective of the center of mass of the galaxy. The drawings are not to scale. As can be seen, the resultant cosmic acceleration varies from point to point along the circular orbit of a particle, whose angle-dependent general form is indicated on the top right. Note that although the gravitational force of attraction is not indicated in the cartoon, it is going to be parallel to the shown acceleration vectors on the right, which are all centripetal (see text). Note that for $r \ll R$, the center-pointing Coriolis force will be $2\omega^2 R$ and it will dominate over the contribution coming from the centrifugal force.

Putting the two cosmic terms together, we obtain the resultant acceleration shown on various points along the orbit in the right panel of Fig. 7, which varies from point to point. The general form of the position-dependent cosmic acceleration is also indicated in the figure. It is important to note that this resultant acceleration, and hence its force, is pointing toward the center, just like gravity. The net force is thus larger than the gravity alone. This is significant because, it affords larger orbital speeds than the usual Newtonian gravitation would allow, rendering the need for any dark matter or other exotic explanations superfluous.

The net universal acceleration is given by

$$a_{cosmic} = 2\omega^2 R + \omega^2 r(2\sin\theta - 1), \tag{22}$$

where the angle $\theta$ runs from 0 to $2\pi$.

Finally, we need to take into consideration the gravitational acceleration, which, for a galaxy of mass $M$, is given by $GM/r^2$ and is also pointing toward the galactic center of mass. Putting all this together in Newton's second law we obtain

$$2\omega^2 R + \omega^2 r(2\sin\theta - 1) + \frac{GM}{r^2} = \frac{v^2}{r}, \tag{23}$$

where $v$ is the orbital velocity of a particle in orbit in the galaxy. Note that for the case of no rotation ($\omega = 0$), we recover the familiar Newtonian scheme.

Using the key postulate of the proposed rotational model, i.e., $\omega = H$ we rewrite Eq. (23) as

$$2H^2 R + H^2 r(2\sin\theta - 1) + \frac{GM}{r^2} = \frac{v^2}{r}, \tag{24}$$

Eq. 24 is, according to my rotary model, the most general form of the dynamics of galactic systems. It shows that the velocity of a particle is dependent not only on its position in the galactic orbit, but also on its position with respect to the cosmic axis. The latter furnishes us with a falsifiable test: the galaxy rotation rates should be position dependent as given in Eq. (24).

However, surveying scores of galaxies observationally, one is dealing with a menagerie of different rotating galaxies with varying orientations, and cosmic radii. Therefore, to get a workable and testable model, we need the average of the cosmic acceleration in Eq. (24) (the first two terms). Noting that the first term changes linearly with $R$, whose the interval is $[0, R_0]$, where $R_0$ is the Hubble radius, and that the angle $\theta$ in the second term varies between $[0, 2\pi]$, we will obtain

$$H^2 R_0 - H^2 r + \frac{GM}{r^2} = \frac{v^2}{r}. \tag{25}$$

Using Hubble's law, we arrive at

$$Hc - H^2 r + \frac{GM}{r^2} = \frac{v^2}{r}. \tag{26}$$

Here, we note that the $Hc - H^2 r$ terms is the universal contribution due to the cosmic Coriolis and centrifugal forces, respectively.

The corresponding potential expression per unit mass for Eq. (25) would be

$$V(r) = +Hcr - \frac{H^2 r^2}{2} - \frac{GM}{r}, \tag{27}$$

where we have, apart from the usual gravitational potential function, one linear term and one quadratic term generated by the cosmic Coriolis and centrifugal forces, respectively, due to the postulated rotation of the universe.

How do these results fare against the observational data of RCs and the leading candidate theories? First we note that using the universal constants in Eq. (26), we get (in units of acceleration)

$$\frac{v^2}{r} = 6.59 \times 10^{-10} - 5.3 \times 10^{-36} r + \frac{GM}{r^2}. \tag{28}$$

Eq. (28) implies that for distances close to the galaxy center, the standard gravity term will dominate, as expected from Newtonian gravity. Then as the gravitational acceleration term gets smaller with distance below the critical value of $\bar{a} = Hc = 6.59 \times 10^{-10} \, m/s^2$, the standard gravity term will be dominated by the first term due to the cosmic Coriolis force, giving rise to anomalies. Finally, for very large distances the second term will take over, which is due to the cosmic centrifugal force.

Therefore, in addition to the aforementioned constraint on the angular velocity of cosmic rotation, a second observational criterion that is implied by my rotational model is the existence and effect of two ubiquitous accelerations (Coriolis and centrifugal) generated by cosmic rotation. This implies that for sufficiently large physical systems (e.g., galaxies, galaxy clusters and superclusters), the effect of these accelerations will be significant, which will compel astronomers, who are unaware of a cosmic rotation, either to postulate the existence of a mysterious substance (e.g., DM) or to modify the Newton-Einstein dynamics to account for the anomaly.

What is noteworthy is that the acceleration in Eq. (28) due to the universal Coriolis force takes the constant value of $6.59 \times 10^{-10}\ m/s^2$. Therefore for distances not too small and not too large, it is this universal term that will dominate the centripetal acceleration of particles in galaxies. Remarkably, such a universal acceleration is a common feature of RCs of multitudes of different galaxies. In fact, the appearance of such a cosmic term—that is, H, the Hubble constant—in the local dynamics of galactic systems has been baffling to astrophysicist.[28,31] Furthermore, such a critical acceleration parameter also appears, within an order of magnitude, in the three leading theories to explain the RC anomaly problem, namely MOND $\left(\frac{6.9}{2\pi} \times 10^{-10}\ m/s^2\right)$, MSTG $(6.9 \times 10^{-10}\ m/s^2)$, and CG $(2.75 \times 10^{-11}\ m/s^2)$.[29] The universality of such an acceleration parameter to observations and these different theories has been hailed as encouraging, although its underlying mechanism has been a mystery for the researchers in the field.[31]

Thus the rotary model explains this otherwise mysterious universal acceleration, and proffers a robust physical basis for the link between the cosmic and the local physics.

To generalize our scheme to apply for different cosmic distances and different galaxies, we parameterize Eq. (27) as follows:

$$V(r)_{rot} = +\alpha(Hc)r - \beta\frac{(Hc)^2 r^2}{2} - \frac{GM}{r},\tag{29}$$

where $\alpha$ and $\beta$ are constants to be determined from fitting the observational data.

To further point out the power and simplicity with which the rotary model reproduces the correct equation for galactic RCs, I draw the reader's attention to the highly successful working formula of the CG theory for the gravitational potential, by Mannheim and coworkers, for explaining RCs of many galaxies.[32] They first detected a universal linear potential term by applying their CG theory to a limited set of RC data. However, after applying their theory to a larger set of RC data that contained galaxies whose data points extended well beyond the optical disks, they identified a second globally induced, de Sitter-like quadratic term. They thus reported the following potential

$$V(r)_{CG} = +\left(\frac{\gamma_0}{2}c^2\right)r - \left(\frac{\kappa}{2}c^2\right)r^2 - \frac{GM}{r},\tag{30}$$

where $\gamma_0$ and $\kappa$ are constants whose values are extracted from data fitting. The corresponding centripetal acceleration formula of Eq. (30) has been used for fitting data from scores of different kinds of galaxies, and has been found to capture the essence of the data without the need for any DM. They concluded that their study "suggests that invoking dark matter may be nothing more than an attempt to describe global physics effects such as these in purely local galactic terms."

Comparing Eqs. (27) and (29) of the rotary model with Eq. (30), one sees the phenomenal agreement in the form of the gravitational potentials, one linear term and one quadratic term, in their signs and dependence on the distance from the galactic center.

Curiously, Mannheim and coworkers attribute the existence of the linear term to the "homogenous cosmic background," and that of the quadratic term to the "inhomogeneities in the cosmic background." Furthermore, they recognize that the latter has a de-Sitter-like (an empty universe model with $\rho = k = p = 0$, but with a positive cosmological constant) form although, they contend, it is not associated with an explicit de Sitter geometry.

As I have shown in Part I, the de Sitter solution is nothing but a holding-onto the rotational energy of the universe even while stripping it out of its material content. I have reproduced here the quadratic de Sitter term from the postulated cosmic centrifugal force, and the linear term from the cosmic Coriolis force. The former contains in it the galactic radius, and hence reflects the local contributions to RCs. On the other hand, the latter does not contain any local parameter in it, and hence reflects only global contributions coming from the universal rotation. So with the rotary model, the "homogeneity" and "inhomogeneity" argument to explain the RC anomalies finds a natural physical justification and meaning.

For a comparison between the observation-based physics of the galaxy rotation curves and the results of my rotary model, refer to Table I.

It pays to note here that the effect of a cosmic Coriolis force is to create vortex on galaxies and galaxy clusters, for its curl is $\nabla \times F_\theta \neq 0$. Therefore, with this model we can also make sense of the spin and the stunning similarity of the shapes of the Earthly hurricanes and celestial galaxies. The former are forged as a result of an interplay between the forces of local pressure gradient and global (Earthly) inertial (Coriolis and centrifugal) effects. Now, the herein presented rotary model contends that the galaxies are also forged in a similar way out of a cosmic interplay between the forces of local (galactic) gravity and cosmic inertial (Coriolis and centrifugal) effects. Hence, the stunning similarity of their morphologies.

Lastly, my model rules out any exotic explanation of the DM problem—through, for example, WIMPs and MACHOs—and explains the lack of interaction of "dark matter" with electromagnetic force, and thus unveils the mysterious nature of "dark matter". On the other hand, my model vindicates the machinery of the nonstandard gravity theories such as MOND, CG and MSTG, but condemns their basic philosophical premise that Newton-Einstein dynamics is faulty at low accelerations.[28,30,31,33]

**Conclusions**

In summary, I have demonstrated that a rotating disk-shaped universe model—with an angular frequency equal to the Hubble constant—resolves some of most outstanding problems in cosmology today in a heartbeat. Specifically, in the first part, I have shown that the rotary model naturally explains many seemingly mysterious and difficult issues surrounding dark energy such as the origin and extreme small value of the cosmological constant/dark energy, its smoothness, its gravitationally repulsive nature and Hubble's law. The model also successfully predicts the recent onset of cosmic acceleration with a high degree of precision.

In the second part, I have shown that dark matter is also a cosmic inertial effect resulting from the rotation of the universe, specifically due to the cosmic Coriolis and centrifugal forces. The hitherto mysterious universal acceleration parameter, and the global contributions in the local dynamics of galaxies—as encountered, but not at all understood, in observations and the observationally-backed nonstandard gravity theories of MOND, CG and MSTG—all find a natural explanation in the rotary model without the additional cost of forsaking the standard Newton-Einstein gravity paradigm. Furthermore, the rotary model naturally produces the linear and quadratic potential terms recently deduced from galactic rotation curves as nothing but the effects of cosmic Coriolis and centrifugal forces, respectively.

My model thus has significant implications for the cosmological principle and the standard model of cosmology. Our universe appears to possess mysterious dark energy and dark matter because it is part of a rotating spacetime of a disk-shaped universe, and the centrifugal and Coriolis forces due to rotation are perceived by us as dark energy and dark matter.

**References**

[1]  S. M. Carroll, Living Rev. Relativity **3**, 1 (2001).

[2]  J. Garriga, A. Vilenkin, Phys. Rev. D. **61**, 083502-1 (2000).

[3]  R. Penrose, *The Road to Reality*, (Vintage Books, New York, 2007).

[4]  P. J. Steinhardt, N. Turok, Science **312**, 1180-1182 (2006).

[5]  R. R. Caldwell, Rahul Dave, and P. J. Steinhardt, Phys. Rev. Lett. **80**, 1582-1585 (1998).

[6]  C. Beck and M. C. Mackey, Int. J. Mod. Phys. **D17**, 71-80 (2008).

[7]  R. R. Caldwell *et al*., 2003, Phys. Rev. Lett. **9**, 071301-1.

[8]  S. M. Carroll, astro-ph/0310342 (2003).

[9]  C. Beck and M. C. Mackey, Int. J. Mod. Phys. **D17**, 71-80 (2008).

[10]  S. Weinberg, Rev. Mod. Phys. **61**, 1 (1989).

[11]  S. Zorba, Mod. Phys. Lett. A, **27**, 1250106 (2012).

[12]  K. Gödel K., Rev. Mod. Phys. **21** (3), 447 (1949).

[13]  S.-C. Su, M.-C Chu, ApJ, 703, 354 (2009).

[14]  R. A. Mould, *Basic Relativity* (Springer-Verlag, New York, NY, 1994).

[15]  O. Bikwa, F. Melia, A. Shevchuk, MNRAS, 000, 1 (2010).

[16]  Tegmark M., *et al.*, Physical Review D **69**, 103501 (2004).

[17]  S. M. Carroll, *Spacetime and Geometry* (Addison Wesley, San Francisco, CA, 2004).

[18]  T.-P. Cheng, *Relativity, Gravitation, and Cosmology* (Oxford Univ. Press, New York, NY, 2005).

[19]  W. Rindler, *Relativity* (Oxford Univ. Press, New York, NY, 2006).

[20]  J. W. Rohlf, *Modern Physics from $\alpha$ to $Z^0$* (Wiley, New York, NY, 1994).

[21]  M. P. Hobson, G. Efstathiou, A. N. Lasenby, *General Relativity* (Cambridge Univ. Press, New York, NY, 2006).

[22]  L. Suskind (2008). *General Relativity* [Video file]. Retrieved from http://academicearth.org/courses/foundations-of-modern-physics.

[23]  L. Anderson *et al.*, arXiv:1203.6594v1 [astro-ph.CO] (2012).

[24]  B. A. Reid *et al.*, arXiv:1203.6641v1 [astro-ph.CO] (2012).

[25]  E. F. Taylor, J. A. Wheeler, *Exploring the Black Holes* (Addison Wesley Longman, 2000).

[26]  P. D. Mannheim, astro-ph/0505266v2 (2005).

[27]  S. Funkhouser, Proc. R. Soc. A **464**, 1345 (2008).

[28]  J. D. Bekenstein, Contemporary Physics **47**, 387 (2006).

[29]  P. D. Mannheim, J. G. O'Brien, arXiv:1011.3495v4 [astro-ph.CO] (2012).

[30]  G. D'Amico, M. Kamionkowksi, K. Sigurdson, arXiv:0907.1912v1 [astro-ph.CO] (2009).

[31]  M. Milgrom, Proceedings of Science PoS(HRMS)033; arXiv:1101.5122v1 (2010).

[32]  P. D. Mannheim, J. G. O'Brien, Phys. Rev. Lett. **106**, 121101 (2011).

[33]  P. D. Mannheim, Prog. Part. Nucl. Phys. **56**, 340 (2006).